\documentstyle[preprint,epsfig,aps]{revtex}
 \tighten
\def\bea{\begin{eqnarray}}
\def\beann{\begin{eqnarray*}}
\def\beq{\begin{equation}}
\def\eea{\end{eqnarray}}
\def\eeann{\end{eqnarray*}}
\def\eeq{\end{equation}}
\def\nn{\nonumber}

\begin{document}
\draft
\title{Short-range repulsion and isospin dependence in the $KN$ system}
\author{
D. Hadjimichef$^{1}$, J. Haidenbauer$^{2}$, and G. Krein$^{3}$ \\
{\small $^1$ Departamento de F\'{\i}sica, Universidade Federal de Pelotas, 
96010-900 Pelotas, RS, Brazil} \\
{\small $^2$ Forschungszentrum J\"ulich, Institut f\"ur Kernphysik, 
D-52425 J\"ulich, Germany} \\
{\small $^3$ Instituto de F\'{\i}sica Te\'{o}rica, Universidade Estadual 
Paulista } \\
{\small Rua Pamplona, 145 - 01405-900 S\~{a}o Paulo, SP, Brazil} 
}
\maketitle
\begin{abstract}
The short-range properties of the $KN$ interaction are studied  
within the meson-exchange model of the J\"ulich group.
Specifically, dynamical explanations for the phenomenological 
short-range repulsion,  
required in this model for achieving agreement with the empirical
$KN$ data, are explored. Evidence is found that contributions from
the exchange of a heavy scalar-isovector meson ($a_0$(980)) as well
as from genuine quark-gluon exchange processes are needed. Taking
both mechanisms into account a satisfactory description of the
$KN$ phase shifts can be obtained without resorting to 
phenomenological pieces. 
\end{abstract}
\vspace{2.5cm}
\noindent{PACS NUMBERS: 13.75.Jz, 12.39.-x, 14.20.Jn, 21.30.-x, 12.40.-y}

\vspace{1.0cm}

\newpage 
\section{Introduction}

The kaon-nucleon ($KN$) system provides an ideal setting for studying
short-distance effects of the hadron-hadron force. This is because 
pions play a much less important role in the $KN$ system than in the 
most extensively studied nucleon-nucleon ($NN$) system. Indeed, the
one-pion-exchange is absent in the $KN$ interaction and the 
contributions from $2\pi$-exchange to the interaction seem to be 
weaker than in the $NN$ system.  Under such circumstances one expects 
that short-distance effects can be most easily isolated from the 
attractive medium-range background and that possibly effects from 
explicit quark-gluon degrees of freedom can be identified.

A large body of work accumulated in the last 50 years indicates that
meson degrees of freedom are very efficient for describing low-energy
hadron-hadron interactions. In particular for the $KN$ system, a few
years ago the J\"ulich group presented a meson-exchange model for
the $K^+N$ scattering~\cite{Juel1,Juel2}. Ref.~\cite{Juel1} considered 
single boson exchanges ($\sigma$, $\rho$, $\omega$), together with 
contributions from higher-order diagrams involving $N$, $\Delta$, $K$ 
and $K^*$ intermediate states. It turned out that the $S$-wave observables 
of $KN$ experiments could only be described with the model if the value of 
the $KK\omega$ coupling constant is increased about 60\% above the value 
that follows from the SU(3) (quark flavor) symmetry. Specifically, the 
increased value of the $KK\omega$ coupling constant was strictly necessary 
to obtain the $S$-wave low-energy parameters and the energy dependence of 
$S$-wave phase shifts for both isospin $I=0$ and $I=1$ channels. However, 
this increased value lead to additional repulsion in the $P$ and higher 
partial waves which seemed to be not favored by the empirical data, especially 
in the $P_{03}$ and $P_{13}$ channels~\footnote{The spectroscopic notation 
used is such that a partial wave with angular momentum $L$, total angular
momentum $J$ and isospin $I$ is denoted by $L_{I\,2J}$.}. 
Thus, it was concluded in Ref.~\cite{Juel1} that the required additional
contributions must be much shorter ranged than the $\omega$ exchange. 

Further evidence for the conjecture that the repulsion needed to 
describe $KN$ scattering cannot be interpreted completely in terms of
conventional $\omega$-exchange came from subsequent investigations of 
the $\bar K N$ system \cite{Mueller}. 
In a meson-exchange model like the one developed by the J\"ulich
group there is a close connection between the $KN$ and $\bar KN$ interactions
due to G-parity conservation. Specifically, this means that the
repulsive $\omega$-exchange changes sign for $K^-N$, because of the
negative G-parity of the $\omega$-meson, and becomes attractive. 
A large contribution from the $\omega$-exchange as favored by the
$KN$ $S$-waves turns then into a strongly attractive piece -- which
is indeed much too strong to fit the $K^-N$ data \cite{Mueller}.

The conclusion from those results was that $\omega$-exchange, as treated 
in this model, can only be interpreted as an effective contribution that 
parameterizes besides the ``physical''
$\omega$-exchange also further shorter-ranged mesonic contributions or
genuine quark-gluon effects or both. 
This was shown by a model analysis where the coupling constants of the
$\omega$ meson ($g_{KK\omega}$, $g_{NN\omega}$) were kept at their 
SU(3) symmetry values and an additional phenomenological
(extremely short-ranged) repulsive contribution, a ``$\sigma_{rep}$'',
with a mass of about 1.2 GeV was added - see Figs.~1(a) and 1(b).  

In Ref.~\cite{Juel2} the model was further refined by replacing 
$\sigma$-- and $\rho$--exchange by the correlated $2\pi$--exchange 
contribution in the $J^P=0^+$ and $J^P=1^-$ channels, respectively,
as illustrated in Fig.~1(c). This was done by starting from a 
microscopic model for the $t$--channel reaction $N \overline N \rightarrow 
K \overline K$ with $\pi\pi$ (and $K \overline K$) intermediate states 
and using a dispersion relation over the unitarity cut. Such a realistic 
model of (effective) $\sigma$-- and $\rho$--exchange was then used to
reconstruct an extended meson exchange model for $KN$ scattering. 
But again, as is Ref.~\cite{Juel1}, the addition of a phenomenological 
$\sigma_{rep}$ was essential to describe the data with
the SU(3) $KK\omega$ coupling constant. 

One possible interpretation for the need of a very short-ranged repulsion,
shorter ranged than that provided by $\omega$-exchange, is that quark-gluon
effects are playing a role~\cite{Juel1}. The study of the $KN$ interaction in the 
context of quark models has a long history since the 1980's~\cite{earlier}. 
More recently, the subject has gained renewed interest with the works of 
Barnes and Swanson~\cite{BS-KN} and Silvestre-Brac and 
collaborators~\cite{SB1,SB2,SB3}. The main ingredients in the calculations of
both groups are the nonrelativistic quark model and a quark interchange 
mechanism with one-gluon-exchange (OGE). One important conclusion of these 
calculations is that the derived $KN$ interaction is short-ranged and repulsive,
and strongly isospin dependent. As we discussed above, although in the J\"ulich 
model the overall strength and energy dependence of the $S$-wave phase-shift of 
$K^+N$ scattering can be obtained by augmenting the value of the $KK\omega$ 
coupling, the $P$ and higher partial waves do not come out right and the 
introduction of the exchange of a fictitious scalar particle with repulsive 
character was essential for this matter. In view of this, the substitution of 
an isospin independent $\sigma_{rep}$ by a strongly isospin-dependent 
quark-gluon dynamics is not trivial and apparently bound to fail. However,
we note that in both Refs.~\cite{Juel1,Juel2} the $a_0(980)$ meson
was left out without apparent reason. This meson, being a scalar-isovector, is 
an important source of isospin dependence and it has a mass not much larger
than those of the other mesons considered in the model. Thus,  
it can, in principle, play an important role for the isospin dependence of
the $KN$ interaction. 

The main motivation of our paper is to investigate the energy and isospin 
dependences of the $KN$ interaction in a hybrid model, in which the J\"ulich 
model is extended by adding the $a_0(980)$ exchange, and where 
the very short-ranged 
part of the $KN$ interaction is described by quark-gluon exchange, instead of 
the phenomenological $\sigma_{rep}$. We construct an effective $KN$ potential
from a microscopic nonrelativistic quark model from the interquark exchange 
mechanism and use all components of the OGE interaction. These include
the Coulomb, spin-independent contact and spin-orbit interactions. 
In addition, we use a linearly rising potential to represent quark 
confinement. 

The paper is organized as follows. In the next section we describe
the $KN$ meson-exchange model of the J\"ulich group. 
In section III we provide an overview of studies on the $KN$
system that were carried out in the framework of the quark-model. 
In section IV we present and discuss our results. Specifically,
we investigate
the consequences of replacing the phenomenological $\sigma_{rep}$ of
the J\"ulich model by quark-gluon exchange.
Our conclusions and perspectives are presented in
Section~\ref{sec:ConPer}. Appendix A presents the expressions for the
effective $KN$ potentials in the quark model.

\section{The J\"ulich $KN$ model}
\label{sec:meson}

The J\"ulich meson-exchange model of the $KN$ interaction has 
been widely described in the literature \cite{Juel1,Juel2,Juel3,Juel4}
and we refer the reader to those works for details. Here we 
will only summarize the features which are relevant for the
present study. 

The J\"ulich meson-exchange model of the $KN$ interaction was 
constructed along
the lines of the (full) Bonn $NN$ model \cite{MHE} and its extension 
to the hyperon-nucleon ($YN$) system \cite{Holz}. 
Specifically, this means that one
has used the same scheme (time-ordered perturbation theory), the same 
type of processes, and vertex parameters (coupling constants, cut-off
masses of the vertex form-factors) fixed already by the study of 
these other reactions. 

The diagrams considered for the $KN$ interaction are shown in
Fig.~\ref{Juel_mod}. Based on these diagrams a $KN$ potential $V$ is
derived, and the corresponding reaction amplitude
$T$ is then obtained by a solving a Lippmann-Schwinger type 
equation defined by time-ordered perturbation theory:
\begin{equation}
T = V + V G_0 T \ .
\label{LSE}
\end{equation}
From this amplitude phase shifts and observables (cross sections,
polarizations) can be obtained in the usual way.

As seen in Fig.~\ref{Juel_mod}, 
obviously the J\"ulich model contains not only single-meson
(and baryon) exchanges, but also higher-order box diagrams 
involving $NK^*$, $\Delta K$ and $\Delta K^*$ intermediate 
states. Most vertex parameters involving the nucleon and the $\Delta$(1232)
isobar can be taken over from the (full) Bonn $NN$ potential. 
The coupling constants at vertices involving strange baryons are fixed 
from the $YN$ model (model B of Ref.~\cite{Holz}). Those quantities
($g_{N\Lambda K}$, $g_{N\Sigma K}$, $g_{NY^* K}$) have been related to the 
empirical $NN\pi$ coupling by the assumption of SU(6) symmetry, 
cf. Ref.~\cite{Juel1,Juel2}.

For the vertices involving mesons only, most coupling constants
have been fixed by SU(3) relating them to the empirical $\rho \to
2\pi$ decay. Exceptions are the coupling constants $g_{KK\sigma}$ 
and $g_{KK\omega}$, which have been adjusted to the $KN$ data, for 
the following reason: The $\sigma$ meson (with a mass of about
600 MeV) is not considered as a genuine particle but as a simple
parametrization of correlated $2\pi$-exchange processes in the 
scalar-isoscalar channel. Therefore, its coupling strength cannot
be taken from symmetry relations.  
Concerning the $\omega$-exchange it was found that a much 
larger strength than obtained from SU(3) was required in order to
obtain sufficient short-range repulsion for a reasonable description of 
the $S$-wave $KN$ phase shifts \cite{Juel1}. The $\omega$-coupling 
$g_{KK\omega}$
had to be increased by about 60\% over the symmetry value - quite
analogous to the situation in the $NN$ system \cite{MHE}. 
However, such an increased $\omega$-exchange turned out to be in
contradiction with the empirical data on $P$- and higher partial
waves. Specifically $P_{03}$ and $P_{13}$ do not really demand 
additional repulsive contributions. Thus, it was concluded that
the additional repulsion should be of rather short-ranged nature.
Such a contribution would still allow to obtain a reasonable
description of the $S$-waves, but would leave $P$- and higher
partial waves basically unchanged.  

As a consequence the J\"ulich group presented a model
where the coupling strengths for both $g_{NN\omega}$ and $g_{KK\omega}$
were kept at their SU(6) values. At the same time a phenomenological,
very short-ranged contribution was added. This phenomenological piece 
has the same analytical form as $\sigma$-exchange, but an exchange mass
of 1200 MeV and, most importantly, an opposite sign. Accordingly, 
it was denoted $\sigma_{rep}$.  

In a subsequent investigation the $\sigma$(600) and also the 
elementary $\rho$ were replaced by a microscopic model for
correlated 2$\pi$ (and $K\bar K$) exchange between kaon and nucleon,
in the corresponding scalar-isoscalar and vector-isovector
channels \cite{Juel2}. Starting point for this was a model
for the reaction $N\bar N \to K\bar K$ with intermediate 2$\pi$
and $K\bar K$ states, based on a transition in terms of baryon
($N$, $\Delta$, $\Lambda$, $\Sigma$) exchange and a realistic
coupled channel $\pi\pi \to \pi\pi$, $\pi\pi \to K\bar K$, and
$K\bar K \to K\bar K$ amplitude. The contribution in the $s$-channel 
is then obtained by performing a dispersion relation over the 
unitarity cut. But also in this model the phenomenological
short-ranged $\sigma_{rep}$ was needed in order to achieve agreement
with the empirical phase shifts. 

Since the results of Ref.~\cite{Juel2} indicate that the 
contributions of the correlated 2$\pi$ exchange in the scalar-isoscalar
channel are in rough agreement with the
effective description by $\sigma$-exchange used in Ref.~\cite{Juel1}
we will employ the latter in the present investigation for simplicity
reasons. Specifically, we will use the $KN$ model I presented in
Ref. \cite{Juel2}. The parameters of this model are 
summarized in Table I.
Resulting phase shifts for the J\"ulich model I \cite{Juel2} will
be shown and compared with empirical data in sect. IV. 
Further results, including also scattering observables can be found 
in Ref.~\cite{Juel2}.

\section{The $KN$ interaction in the quark model}
\label{sec:sigma}

In this section we briefly review the salient features of 
model calculations of the $KN$ interaction that are based on
quark-gluon exchange and derived within the nonrelativistic quark model. 
Specifically, we will focus on the more recent calculations of 
Barnes and Swanson~\cite{BS-KN} and Silvestre-Brac and 
collaborators~\cite{SB1,SB2,SB3}.

Barnes and Swanson use the quark-Born-diagram (QBD) method~\cite{QBD}.
In this method, the $KN$ scattering amplitude is assumed to be the
coherent sum of all one-gluon-exchange (OGE) interactions followed by 
all allowed quark line exchanges. The input to this method is the
microscopic quark-quark interaction and kaon and nucleon wave 
functions. Barnes and Swanson~\cite{BS-KN} used the contact spin-spin 
``color-hyperfine'' component of the OGE and used Gaussian wave 
functions for the interacting hadrons, which allowed them to evaluate 
the $KN$ scattering amplitude analytically. They calculated isospin 
$I=0$ and $I=1$ scattering observables such as $S$-wave phase shifts 
and scattering lengths. The model has only two parameters, the ratio 
of the $u,d$ to $s$ quark masses, $\rho = m_q/m_s$, and 
$\alpha_s/m_q^2$, where $\alpha_s$ is the quark-gluon coupling 
constant. By using typical quark-model parameters and using the 
Born approximation, Barnes and Swanson obtained very reasonable 
results for the $S$-wave phase shifts~\cite{BS-KN}. 

The studies of Silvestre-Brack and collaborators~\cite{SB1,SB2,SB3} 
complement the work of Barnes and Swanson in several aspects. First 
of all, Refs.~\cite{SB1,SB2,SB3} employ the RGM instead of the QBD 
method. The main difference between the two methods refers to the 
way orthogonality effects in the relative $KN$ wave functions are 
treated. These are effects due to the Pauli principle for quarks
in different clusters. It can be shown~\cite{FT} that the differences
between effective hadron-hadron interactions calculated with both 
methods are usually small, although not entirely negligible. In
Ref.~\cite{SB1} the authors calculate $S$-wave scattering phase shifts
using the Coulomb, spin-spin and constant contact pieces of the OGE 
quark-quark interaction. In addition, they include the exchange of $\pi$ 
and $\sigma$ mesons - considered as elementary particles - between quarks, 
and a linearly rising confining potential. The same quark-quark interaction 
is used to build the $K$ and $N$ wave functions and to generate the $KN$
interaction. The parameters are constrained to reproduce the low-lying meson 
and baryon spectrum. The main conclusion of this study was that it is 
impossible to describe both $I=0$ and $I=1$ isospin channels simultaneously 
within the model. Relativistic kinetic energy effects were investigated 
Ref.~\cite{SB2}. The results obtained for the $S$-waves, for which one 
expects such effects to be more important, were not much different from 
the corresponding nonrelativistic ones. In Ref.~\cite{SB3}, scattering 
phase shifts from $S$- up to $G$-waves were calculated. The difference from 
Ref.~\cite{SB1} is that a spin-orbit interaction was added to the OGE and 
confining pieces used there. No meson exchanges between quarks were considered.
The parameters were again fixed by requiring a good description of the 
low-lying meson and baryon spectrum. The results obtained are such that
the $I=0$ $S$-wave phase shift is reasonably well described, while the 
corresponding $I=1$ is too repulsive. The $P$ and higher partial-wave phase
shifts are poorly described, with the exception of the $P_{11}$, $D_{13}$,
$D_{15}$, and $G_{19}$ phases. The $P_{01}$ phase, for example, is predicted 
to be almost zero, while the corresponding experimental phase grows from 
zero up to 60 degrees at $p_{lab} = 1$~GeV. 

The results of Silvestre-Brac and collaborators~\cite{SB1,SB2,SB3} clearly
indicate that the quark-interchange mechanism with OGE alone is not sufficient 
to describe the $K^+N$ data.
However, it seems to provide at least enough strength for $S$-waves. 
In view of the discussion above on the J\"ulich model, we investigate 
here the substitution of $\sigma_{rep}$ in that model by the quark-interchange
mechanism with OGE. Recently, one of us~\cite{Dimi} has derived the 
contribution of the spin-spin part of the OGE to the central part of the $KN$ 
interaction using the mapping formalism developed in Ref.~\cite{FT}. The 
different contributions to the $K^+N$ effective potential from the 
quark-interchange mechanism are illustrated in Fig.~\ref{FT-KN}. 
The on-shell $KN$ amplitude is identical to the one derived by Barnes and 
Swanson~\cite{BS-KN}. In order to iterate the potential in a Lippmann-Schwinger
equation one needs the off-shell amplitude. For this purpose, we have calculated 
the contributions of all remaining components of the OGE to the $K^+N$ effective 
potential within the framework of Ref.~\cite{FT}. We found that the spin-spin
component of the OGE gives by far the most important contribution to the $K^+N$
effective potential. 

For illustrative purposes we present, in this section, the phases calculated in 
Born approximation - in the next section the OGE $KN$ potential is iterated in
the Lippmann-Schwinger type of equation, Eq.~(\ref{LSE}). 
In Fig.~\ref{phasqf} we present the $S$- 
and $P$-wave phase shifts resulting from the OGE and the confining interaction. 
The higher partial waves are very small and are not shown. The experimental 
data points in this figure are taken from Refs.~\cite{Exp1,Exp2,Exp3}.
The analytical expressions for the $KN$ potential are given in the Appendix. 
The parameters of the potential are the masses of the constituent quarks, 
$m_q (= m_u = m_d)$ and $m_s$, the quark-gluon hyperfine coupling 
$\alpha_s$, and the size parameters of the nucleon and kaon wave functions, 
$\alpha$ and $\beta$. We use the ``reference parameter set'' of Barnes and 
Swanson~\cite{BS-KN}, which are conventional quark model parameters.
These are
\bea
\rho &=& m_q/m_s = 0.33 {\,\rm GeV} / 0.55 {\,\rm GeV} = 0.6 \nn \\
\alpha_s/m^2_q &=& 0.6 / (0.33)^2 {\,\rm GeV}^2 \nn \\
\alpha &=& 0.4 {\,\rm GeV} \hspace{1.0cm} \beta = 0.35 {\,\rm GeV} .
\eea
The string tension of the confining potential is taken to be 
$\sigma = 0.18~{\rm GeV}^2$~\cite{black}. In addition, when calculating 
the phase shifts we use the physical masses of the nucleon and the kaon, 
$M_N = 0.940~{\,\rm GeV}$ and $M_K = 0.495~{\,\rm GeV}$. Fig.~\ref{phasqf} 
shows that $S$-waves are reasonably well described by the model, although 
the $I=0$ phase agrees less
well with the data at higher energies. The fit can be improved slightly by 
choosing another set of parameters, as done by Barnes and Swanson in their 
study of the $S$-wave phase shifts~\cite{BS-KN}. In this paper we maintain the 
reference set, since the general trend of 
the higher partial waves will not be modified by a change of the quark
model parameters.

In Fig.~\ref{phasq} we show the separate contributions of the OGE and confining
interaction to $S$ and $P$ phases. The dominance of the spin-spin component of 
OGE is clearly seen. The confining interaction gives a very small contribution
to all waves, and the only noticeable effect from the other components of the 
OGE is the one from the Coulomb part in the $S_{11}$ wave.

\section{Results and discussion}
\label{sec:a_0}

As discussed in the last two sections, the $KN$ interaction
in the low-energy region ($p_{lab} \le 1$ GeV/c) can be well
understood within the meson-exchange picture. However, a
good quantitative overall description of the data can only be
achieved by adding a phenomenological contribution that is 
extremely short-ranged and repulsive - and therefore affects
essentially the S-waves only. At the same time interaction
models based on quark-gluon degrees of freedom yield only
a mediocre overall reproduction of the $KN$ phase shifts. 
However, the predicted S-waves are in fairly good 
agreement with empirical results suggesting that at least
the short-ranged part of the $KN$ interaction is well
accounted for by the one-gluon exchange mechanism that
is the dominant ingredient in those quark models. It is
therefore tempting to combine the contributions of those two 
complementary approaches to the $KN$ force. Indeed 
such a procedure is in the spirit of the
original $KN$ model of the J\"ulich group where it was
suggested that the short-ranged phenomenological piece added
in this model might be an effective parametrization of
either further short-ranged mesonic contributions or 
of genuine quark-gluon effects or both \cite{Juel1,Holi}. 

Results for the $KN$ phase shifts of the original
J\"ulich model (note that we use here model I of Ref. \cite{Juel2}) 
are shown by the dash-dotted lines in Fig.~\ref{phases}. If we switch
off the contribution from the phenomenological $\sigma_{rep}$
and add the contribution from one-gluon exchange instead  
we obtain the short dashed lines. The parameters of the quark model
are the same as in the previous section.
We see that the OGE is indeed capable of producing repulsive
contributions which are of a comparable order of magnitude as the 
one of the phenomenological $\sigma_{rep}$. Indeed, after the
discussion in sect. III this could have been expected. 
However, it is definitely surprising that for the $S_{11}$
partial wave the results with OGE (and without $\sigma_{rep}$)
are almost identical to the ones of the original J\"ulich model.
In case of the $S_{01}$ the situation is somewhat less
satisfying. Here the repulsion provided by the OGE is significantly
smaller than the one parametrized by the $\sigma_{rep}$. 
This is simply a consequence of the isospin dependence inherent
in the OGE -- the phenomenological $\sigma_{rep}$ is, of 
course, per construction an isoscalar. 
The higher partial waves (in the $I=0$ as well as the $I=1$)
are again only marginally changed as compared to the original
results - testifying that also the OGE is of rather short range. 

The above results can be seen as an indication that the OGE is not 
the only short-range physics that is parametrized by the 
$\sigma_{rep}$ of the J\"ulich $KN$ model. (Indeed one might
argue that this could have been already guessed from the 
difference in the isospin structure!) 
Besides possible higher-order contributions resulting from 
quark-gluon dynamics one should not forget to take into
consideration also further shorter-ranged mesonic contributions. 
Indeed the exchange of the $a_0(980)$ meson, which is a
scalar-isovector particle, is a natural candidate for this. 
With its mass of about 1 GeV its contributions are definitely
of short-ranged nature as required. Furthermore, its isospin
structure leads to attractive contributions in the $I=1$
channel but to repulsive contributions in the $I=0$ channel. 
Thus, it complements the isospin dependence of the OGE in an 
almost ideal way and in conjunction with the latter would lead
to contributions that are almost isospin independent -- as
those of the phenomenological $\sigma_{rep}$. 
The $a_0(980)$ meson is taken into account in the Bonn $NN$
model \cite{MHE} (it is denoted as $\delta$ meson there). However,
for unexplained reasons, it was not included in the original
J\"ulich $KN$ model. 
Subsequent investigations of the J\"ulich group on the structure
of the $a_0(980)$ meson suggested that this resonance can
be understood in terms of strong correlations in the $\pi\eta
- K\bar K$ channel \cite{Janssen}. Thus, the situation is similar to the 
strong $\pi\pi - K\bar K$ correlations in the scalar-isoscalar channel
that are usually effectively parametrized by the $\sigma$ meson. Accordingly, 
it is in the spirit of the Bonn/J\"ulich models of hadronic
reactions to consider the contributions of the $a_0(980)$ meson,
and indeed it is included in the more recently developed models
of the $\pi N$ \cite{Schuetz,Krehl} and hyperon-nucleon ($YN$) \cite{YN}
interactions. 
Following the arguments in Refs.~\cite{Janssen,Schuetz} we do not view 
the $a_0$ exchange as a genuine meson exchange but rather as a 
parametrization of correlations in the mesonic systems in the
scalar-isovector channel. Therefore we consider the $a_0$
coupling constants as free parameters. In principle, $g_{K\bar K a_0}$
could be determined from the $a_0$ decay width into the $K\bar K$
system. However, the experimental information on this quantity is
still very poor, cf. Ref.~\cite{PDG}, and thus cannot provide more than 
a guideline. Note that in the $a_0$ exchange the product of the
$NNa_0$ and $K\bar K a_0$ coupling constants appear, and therefore
we also list only this product in Table I. 

Results including now $a_0$ meson-exchange as well as the OGE 
contributions are shown by the long-dashed lines in Fig.~\ref{phases}. 
The coupling strength of the $a_0$ exchange has been chosen in
such a way that the model prediction for the $S_{01}$ partial
wave agrees roughly with the result of the original J\"ulich $KN$
model. As can be seen in Fig.~\ref{phases}, the inclusion of
the $a_0$ exchange influences also the $KN$ $P$-waves in the $I=0$ 
channel, i.e. the $P_{01}$ and $P_{03}$, whereas all other
higher partial waves remain basically unchanged. As a matter of fact,
the present model based on $a_0$ meson exchange and OGE contributions
yields pretty much the same results as the original J\"ulich model 
utilizing the phenomenological $\sigma_{rep}$. Minor differences occur
only in the $S_{11}$ partial wave, which turns out to be now 
somewhat too less repulsive in comparison to the data. Thus, as
a last step, we have slightly re-adjusted the parameters of the
$\sigma$ meson, cf. Table I, which then leads to the
final results shown by the solid lines in Fig.~\ref{phases}. 
Those results provide clear evidence that a comparable quantitative 
description of the $KN$ interaction can be achieved with a model
that avoids phenomenological contributions like the $\sigma_{rep}$
of the original J\"ulich model. 

Finally, we want to address the question whether a description 
of the $KN$ interaction is possible within the J\"ulich model
without introducing explicit contributions from the OGE. 
Treating the coupling constants of the $\sigma$ and $a_0$ mesons
as completely free parameters we were indeed able to obtain
a reasonable reproduction of the $S_{01}$ and $S_{11}$ partial
waves. However, it could only be achieved by assuming that 
the $\sigma$-exchange contribution is basically zero. Of course,
this is completely unrealistic in view of the results obtained for 
the strength of the correlated two-pion exchange in the $\sigma$ 
channel in Ref.~\cite{Juel2}. Moreover, the description of
the higher partial waves deteriorates significantly if the 
$\sigma$-exchange contribution is so strongly reduced.

\section{Conclusions and Perspectives}
\label{sec:ConPer}

In this paper we have studied the short-range properties of the 
$KN$ interaction. In particular, we have taken the $KN$ meson-exchange 
model of the J\"ulich group and we explored possible dynamical
explanations for a phenomenological (extremely short-ranged) repulsive 
contribution, a ``$\sigma_{rep}$'' with a mass of about 1.2 GeV, 
that is present in the J\"ulich model. Such a phenomenological, repulsive 
and rather short-ranged piece had to be introduced in that model for 
achieving agreement with the empirical $KN$ data.

The very short-ranged nature of this repulsion could be a sign 
that quark-gluon dynamics is playing a role. Therefore, we have
calculated corresponding contributions to the $KN$ interaction 
based on the nonrelativistic quark model and a quark interchange 
mechanism with one-gluon-exchange. 
It turned out that those processes are indeed short-ranged 
and repulsive. However, unlike the phenomenological ``$\sigma_{rep}$'' in
the J\"ulich model, they are also strongly isospin dependent. 
Thus, one-gluon-exchange alone can certainly not explain the 
required short-range physics. 
Consequently, we examined additional short-range physics
that arises in the mesonic sector, and specifically the exchange of
the (scalar-isovector) $a_0(980)$ meson.
Its contribution was not included in the original J\"ulich $KN$ model. 

Due to its isospin structure the $a_0(980)$ exchange provides
attraction in the $I=1$ channel and repulsion in the $I=0$ channel, 
and therefore counterbalances the isospin dependence of the one-gluon
exchange. Taking both mechanisms ($a_0$- as well as one-gluon exchange)
into account yields a short-ranged
and repulsive but basically isospin-independent interaction 
-- similar to the one parametrized by the $\sigma_{rep}$ -- 
and, consequently, a satisfactory description of the
$KN$ phase shifts can be obtained without resorting to 
phenomenological pieces, as demonstrated in the present paper. 

The authors of the original J\"ulich $KN$ model conjectured that
the introduced phenomenological $\sigma_{rep}$ might be an
effective parametrization of either further short-ranged mesonic 
contributions or genuine quark-gluon effects or both \cite{Juel1}.
Our investigation provides strong evidence that the third 
alternative is realized. Specifically, it lends support to the
supposition that effects from quark-gluon degrees of freedom
can be explicitly seen in the $KN$ system. 
Still one can raise the question
whether contributions from genuine quark-gluon dynamics are
really needed. E.g., couldn't their role be taken over by the
exchange of heavier vector mesons, say? To answer this question
it will be very instructive to study the $\bar K N$ system again,
using the present model. For example, the investigations in
Ref.~\cite{Mueller} have shown that the $\bar K N$ data require
only a strongly reduced short-ranged repulsive piece, i.e.
only about 20\% of the phenomenological $\sigma_{rep}$ used in
the $KN$ system. It will be interesting to see whether the
present scenario of combined $a_0$(980) exchange and 
quark-gluon dynamics is able to generate these properties
when going over to the $\bar K N$ system. 
One should note, however, that the treatment of the 
$\bar K N$ channel in the nonrelativistic quark model is more complicated
than the $KN$ system 
since it involves $s$-channel gluon exchange. Special care must be taken
with such processes because it is not clear that the use of perturbative 
massless gluons makes physical sense in this model. Contributions of 
intermediate hybrid $q\bar q g$ states to the process must certainly be
considered. Still the extension of the quark interchange model to incorporate
gluon annihilation in the $\bar KN$ system would be a very interesting new
development.
Investigations along this line are planned for the future.

\acknowledgments{The authors are thankful to Ted Barnes for illuminating
discussions. Financial support for this work was provided in 
part by the international exchange program DLR (Germany, BRA W0B 2F) 
- CNPq (Brazil, 910133/94-8).

\appendix

\section{Contributions to the KN interaction from OGE and confinement}

In this Appendix we present the different contributions
of the OGE and of the confining potential to the $KN$ potential used in the 
present paper. Let's denote the initial and final three-momenta of the 
interacting quarks by $\vec k_1$, $\vec k_2$, $\vec k'_1$ and $\vec k'_2$ 
(prime means final state). It is convenient to define the following 
combinations of momenta $\vec q = \vec k'_1 - \vec k_1 = \vec k_2 - \vec k'_2$, 
$p_1 = (\vec k_1 + \vec k'_1)/2$ and $p_2 = (\vec k_2 + \vec k'_2)/2$. In terms
of these, the interquark interaction can be written as
\beq
H_{OGE} = \sum_{ij}\Bigl[\sum_{a} {\cal F}^a(i){\cal F}^a(j)\Bigr]\,
V_{ij}(\vec q, \vec p_i, \vec p_j) ,
\eeq
where $i,j$ identify the quarks (or antiquarks) $1,2$ and 
$V_{ij}(\vec q, \vec p_i, \vec p_j)$ depends on spin variables and the
indicated momenta. The color SU(3) matrices ${\cal F}^a(i)$, $a=1,\cdots 8$,
are given in terms of the Gell-Mann matrices $\lambda^a$ as 
${\cal F}^a(i) = \lambda^a/2$ when $i$ is a quark and ${\cal F}^a(i) = 
- \lambda^{a T}/2$ when $i$ is an antiquark ($T$ means transpose). 
We refer the reader to the literature for the explicit expression of 
$V_{ij}(\vec q, \vec p_i, \vec p_j)$  - see for example, Eq.~(3) of 
Ref.~\cite{black}. 

Next, we present the individual contributions in 
$V_{ij}(\vec q, \vec p_i, \vec p_j)$ to the $KN$ potential. We represent each
contribution to the $KN$ potential as
\beq
V(\vec p, \vec p\,') = \frac{1}{2} \, \sum^{d}_{D=a} 
\left[V_D(\vec p, \vec p\,') + V_D(\vec p\,', \vec p)\right], 
\label{VKN}
\eeq
where $\vec p$ and $\vec p\,'$ are the initial and final c.m. momenta of
the the $KN$ system, and the index $D$ identifies the diagrams $a, \cdots d$ 
of Fig.~\ref{FT-KN}. The explicit evaluation of these diagrams requires
also the specification of the nucleon and kaon wave functions, $\Psi_N$ and
$\Psi_K$. These are taken to be in momentum space of the form
\beq
\Psi_N(\vec p) = \delta (\vec p - \vec k_1 - \vec k_2 - \vec k_3) \, N(\vec p) 
\, \phi(\vec k_1)\phi(\vec k_2)\phi(\vec k_3) ,
\eeq
where
\beq
\phi(\vec k) = \left(\frac{1}{\pi \alpha^2}\right)^{3/4} 
\exp \left(- {\vec k\,}^2 / 2 \alpha^2  \right) \hspace{1.5cm}
N(\vec p) = \left(3\pi \alpha^2 \right)^{3/4} 
\exp \left( {\vec p\,}^2 / 6 \alpha^2 \right) ,
\label{phi_and N}
\eeq
and
\beq
\Psi_K(\vec p) = \delta (\vec p - \vec k_q - \vec k_{\bar q}) \, 
\left( \frac{1}{\pi \beta^2}\right)^{3/4} \exp \left[ 
-\frac{\left(  m_1\vec k_q - m_2\vec k_{\bar q} \right)^2}
{8\,\beta ^2}\right] , 
\eeq
with 
\bea
m_1=\frac{2\,m_{\bar{q}}}{m_q+m_{\bar{q}}} \hspace{1.5cm}
m_2=\frac{2\,m_q}{m_q+m_{\bar{q}}} .
\eea
For convenience we also introduce the quantities
\beq
\rho = m_q / m_s \hspace{1.5cm} g^2 = \alpha/\beta \hspace{1.5cm} b = 1/\alpha .
\eeq

The explicit contributions of the different pieces of the OGE and of the confining
potential to the effective $KN$ interaction are given by the following expressions.

\subsection{Coulomb} 

\beq
V^{Coul}_D(\vec p, \vec p\,') = 4\pi\alpha_s \, \omega_D(I)  
\int_{0}^{\infty }\,ds\,
\eta_{D}(s) \, \exp \left[ -A_{D}(s)\,p^{2} - B_{D}(s)\,p\,'^{\,2}
+ C_{D}(s)\, \vec{p} \cdot \vec{p}\,'\right],
\label{genCoul}
\eeq
where the variable $s$ comes in because we have chosen to perform
a Laplace transform 
of the Coulomb potential $1/q^2$ in order to integrate over the variable $q$, 
and the
coefficients $\omega_D(I)$ ($I$ identifies isospin $I=1$ or $I=0$) come
from summing over color-spin-flavor of quarks. The 
functions $A_D, B_D, C_D$, and $\eta_D$ for each diagram can be written as a
ratio $A_D = n(A_D)/d(A_D)$, $\cdots$, $\eta_D = n(\eta_D)/d(\eta_D)$. 

\vspace{0.25cm}
\noindent
\underline{Diagram (a):}

\bea
n(A_{a}) &=&  24\,b^{4}\,{\beta }^{2}+9\,b^{2}\,m_{1}^{2}
+ (80\,b^{2}\,{\beta }^{2}+24\,b^{4}\,{\beta }^{4}-48\,b^{2}\,
\beta^2 \, m_{1}+6\,m_{1}^{2}+18\,b^{2}\,{\beta }^{2}\,m_{1}^{2} ) \,s
\nn\\
n(B_{a}) &=& n(A_{a}) 
\nn\\ 
n(C_{a}) &=& 8\,b^{4}\,{\beta }^{2}+3\,b^{2}\,m_{1}^{2}
+ (-16\,b^{2}\,{\beta }^{2}+8\,b^{4}\,{\beta }^{4}+16\,b^{2}\,
{\beta }^{2}\,m_{1}+2\,m_{1}^{2} ) \,s 
\nn \\
n(\eta_{a}) &=& 3\,\sqrt{3}\,b^3 / 8\,{\pi}^3 
\nn \\
d(A_{a}) &=& d(B_{a}) = 6 \, d(C_{a}) = 
48 \beta^2 [3\,b^{2} + ( 2+3\,b^{2}\,{\beta }^{2} ) 
\,s ] 
\nn\\
d(\eta_{a}) &=& [3\,b^{2} + ( 2+3\,b^{2}\,{\beta }^{2} ) 
\,s ]^{3/2} 
\nn \\  
\omega_a (1) &=& - 4/9 \hspace{1.0cm} \omega_a(0) = 0.
\eea

\vspace{0.25cm}
\noindent
\underline{Diagram (b):}

\bea
n(A_{b}) &=& - 24 - 80\,b^{2}\,\beta ^{2}+24\,b^{4}\,\beta
^{4}+12\,m_{1}+60\,b^{2}\,\beta ^{2}\,m_{1}-9\,b^{2}\,\beta ^{2}\,m_{1}^{2} 
\nn \\
&-& ( 96\,\beta^{2}+176\,b^{2}\,\beta ^{4}-48\,b^{4}\,\beta^{6}-48\,
\beta ^{2}\,m_{1} 
- 6\,\beta^{2}\,m_{1}^{2} + 120\,b^{2}\,\beta ^{4}\,m_{1}
- 18\,b^{2}\,\beta ^{4}\,m_{1}^{2} ) \,s
\nn \\
n(B_b) &=& 12 + 20\,b^{2}\,\beta^{2} - 24\,b^{4}\,\beta^{4} - 
6\,m_{1}-18\,b^{2}\,\beta^{2}\,m_{1} \nn \\
&+& ( 48\,\beta^{2} + 8\,b^{2}\,\beta ^{4} - 48\,b^{4}\,\beta^{6} 
- 24\,\beta^{2}\,m_{1} - 12\,b^{2}\,\beta^{4}\,m_{1} 
- 3\,\beta^{2}\,m_{1}^{2} - 9\,b^{2}\,\beta^{4}\,m_{1}^{2} ) \,s
\nn \\
n(C_b) &=& 2\,b^{4}\,\beta^{2} + 2\,b^{2}\,(  m_{1} - 1  ) 
+ (  4\,b^{4}\,\beta ^{4}-8\,b^{2}\,\beta ^{2}+8\,b^{2}\,\beta^{2}
\,m_{1}+m_{1}^{2} ) \,s 
\nn \\ 
d(A_b) &=& 2\, d(B_b) = 12\, \beta^2 \, d(C_b) = 24\,\beta^2 \, 
[ 1 + 3\,b^{2}\,\beta ^{2} + (  4\,\beta^{2}\, 
+ 6\,b^{2}\,\beta^{4} ) s ] 
\nn \\
\omega_b(1) &=& 4/9 \hspace{1.5cm} \omega_b(0) = 0
\eea

\vspace{0.25cm}
\noindent
\underline{Diagram (c):}

\bea
A_{c}(s) &=& 64\,b^{4}\,\beta ^{2}+12\,b^{6}\,\beta
^{4}-60\,b^{4}\,\beta ^{2}\,m_{1}+21\,b^{2}\,m_{1}^{2}+36\,b^{4}\,\beta
^{2}\,m_{1}^{2}
\nn \\
&+& (  320\,b^{2}\,\beta ^{2} + 96\,b^{4}\,\beta^{4} 
- 192\,b^{2}\,\beta^{2}\,m_{1} + 24\,m_{1}^{2} 
+ 72\,b^{2}\,\beta^{2}\,m_{1}^{2} ) \,s 
\nn \\
B_{c}(s) &=& 256\,b^{4}\,\beta^{2} + 12\,b^{6}\,\beta^{4} 
- 132\,b^{4}\,\beta^{2}\,m_{1} + 21\,b^{2}\,m_{1}^{2} 
+ 36\,b^{4}\,\beta^{2}\,m_{1}^{2}
\nn \\
&+& (  320\,b^{2}\,\beta^{2} + 96\,b^{4}\,\beta^{4} 
- 192\,b^{2}\,\beta^{2}\,m_{1}+24\,m_{1}^{2} 
+ 72\,b^{2}\,\beta^{2}\,m_{1}^{2} ) \,s
\nn \\
C_{c}(s) &=& - 32\,b^{4}\,\beta^{2} + 4\,b^{6}\,\beta^{4} 
+ 32\,b^{4}\,\beta^{2}\,m_{1} + 7\,b^{2}\,m_{1}^{2} \nn \\
&+&  [ 32\,b^{4}\,\beta^{4} + 64\,b^{2}\,\beta^{2}\,( m_{1}-1 )
+ 8\,m_{1}^{2} ] \,s 
\nn \\
\eta_{c}(s)&=& 24 \sqrt{3}\,b^3 /\pi^3 
\nn \\
d(A_c) &=& d(A_b) = 6 \, d(A_c) = 48\,\beta ^{2} \, 
[ 7\,b^{2} + 6\,b^{4}\, \beta^{2} 
+ (  8+12\,b^{2}\,\beta^{2} ) \,s ]  
\nn \\
d(\eta_c) &=& [ 7\,b^{2} + 6\,b^{4}\, \beta^{2} 
+ (  8 + 12\,b^{2}\,\beta^{2} ) \,s ]^{3/2}
\nn\\
\omega_c(1) &=& 4/9 \hspace{1.5cm} \omega_c(0) = 0
\eea

\vspace{0.25cm}
\noindent
\underline{Diagram (d):}

\bea
n(A_d) &=& 80\,b^{2} + 160\,b^{4}\,\beta ^{2} + 12\,b^{6}\,\beta^{4}
- 72\,b^{2}\,m_{1} - 132\,b^{4}\,\beta^{2}\,m_{1} + 21\,b^{2}\,m_{1}^{2}
+ 36\,b^{4}\,\beta^{2}\,m_{1}^{2} 
\nn \\
&+& (  320\,b^{2}\,\beta ^{2} + 96\,b^{4}\,\beta^{4} 
- 192\,b^{2}\,\beta^{2}\,m_{1} + 24\,m_{1}^{2}
+ 72\,b^{2}\,\beta^{2}\,m_{1}^{2} ) \,s
\nn\\
n(B_d) &=& n(A_d) \nn \\
n(C_d) &=& - 16\,b^{2} - 32\,b^{4}\,\beta^{2} 
+ 4\,b^{6}\,\beta^{4} + 8\,b^{2}\,m_{1} + 20\,b^{4}\,\beta^{2}\,m_{1}
+ b^{2}\,m_{1}^{2}
\nn \\
&+& (  -64\,b^{2}\,\beta^{2} + 32\,b^{4}\,\beta^{4} 
+ 64\,b^{2}\,\beta ^{2}\,m_{1} + 8\,m_{1}^{2} ) \,s
\nn\\
n(\eta_d) &=& -24 \sqrt{3}\,b^3 \beta^3 /\pi^3 
\nn\\
d(A_d) &=& d(B_d) = 6 \, d(C_d) = 48\,(  2+3\,b^{2}\,\beta^{2} )\,
(  1+2\,b^{2}\,\beta^{2} + 4\,\beta^{2}\,s )
\nn \\
d(\eta_d) &=& [ (  2+3\,b^{2}\,\beta^{2} )\,
(  1+2\,b^{2}\,\beta^{2} + 4\,\beta^{2}\,s ) ]^{3/2}
\nn \\
\nn \\
\omega_d(1) &=& - 4/9 \hspace{1.5cm} \omega_d(0) = 0
\eea

\subsection{Spin-orbit} 

\beq
V^{SO}_D (\vec{p},\vec{p}\,') = i (\vec{p}\times \vec{p}\,')\cdot \vec S_N\,
w^{SO}_D(\vec{p},\vec{p}\,'),
\label{vso-kn}
\eeq
with $S_N$ the spin operator of the nucleon and
\beq
w^{SO}_D(\vec{p},\vec{p}\,') = 4 \pi \alpha_s \, \omega_D(I) 
\int_{0}^{\infty }\,ds\, \eta_D(s)\, \exp \left[ -A_{D}(s)\,p^{2} 
- B_{D}(s)\,p\,'^{\,2} + C_{D}(s)\, \vec{p} \cdot \vec{p}\,'\right] ,
\label{wso}
\eeq
where the functions $A, B, C$ are the same as in the Coulomb potential, and the 
$\eta_D$'s are given 
by 
\bea
\eta_a(s) &=&  \frac{3 \,\sqrt{3}\,b^{5}(8-3m_{1})}{32\pi^{3}
\left[ 3\,b^{2}+\left(2 + 3\,b^{2}\,\beta^{2}\right) \,s\right] ^{5/2}} 
\nn\\
\eta_b(s) &=& - \frac{3}{8\pi^{3}} \,\sqrt{\frac{3}{2}}
\frac{b^{3}\,\beta ^{3}\,\left( m_{1}-2\right) \,\left( 4\,b^{2}\,\beta^{2}
+ m_{1}\right) }{\,\left[ 1+3\,b^{2}\,\beta ^{2}+\left( 4\,\beta^{2} + 
6\,b^{2}\,\beta^{4}\right) \,s\right]^{5/2}}
\nn \\
\eta_c(s) &=&  - \frac{3\,\sqrt{3}
\,b^{5}\,\left( \ m_{1}-4\ -2\,b^{2}\,\beta ^{2}\right) }
{\pi^3\left[
7\,b^{2}+6\,b^{4}\,\beta ^{2}+\left( 8\,+12\,b^{2}\,\beta ^{2}\,\right) s%
\right]^{5/2}} \nn \\
\eta^{(1)}_d(s) &=& - \frac{3\sqrt{3}\,b^{4}\,\beta^{4}\,
\left( 1+2\,b^{2}\,\beta ^{2}\right) \,\left( 2\,b^{2}\,\beta^{2} 
+ m_{1}\right) \,\left( 28\,b^{4}\,\beta ^{4}+m_{1}^{2}+8\,b^{2}\,\beta^{2} 
+ 8\,b^{2}\,\beta ^{2}\,m_{1}\right) }{2\pi^3\left( 2+3\,b^{2}\,\beta^{2}\right) 
\,\left( 1+4\,b^{2}\,\beta ^{2} 
+ 3\,b^{4}\,\beta^{4}\right)^{3/2}\,\left( 4+2\,b^{2}\,\beta ^{2}-m_{1}\right) \,
\left[ 1+2\,\beta^{2}\,\left( b^{2}+2\,s\right) \right]^{5/2} } 
\nn \\
\eta^{(2)}_d(s) &=&  \frac{3\sqrt{3}}{2\pi^{3}} 
\frac{b^3\,\beta^{3}\,\left(2 \,b^2 \, \beta^2 + m_1\right) \, 
\left( 28\,b^{4}\,\beta^{4} + m_{1}^{2} + 8\,b^{2}\,\beta^{2} + 
8 b^{2}\,\beta^{2} m_{1}\right)}
{
\left(2 + 3\,b^{2}\,\beta^{2}\right)^{5/2} \,\left( 4\,b^{2}\,
\beta^{2} + m_{1}\right) \, \left[1 + 2\,\beta^{2}\,\left(b^{2} + 2\,s\right) 
\right]^{5/2} }
\nn\\
\omega_a (0) &=& - \frac{4}{3}\, \frac{1}{m^2} \hspace{1.0cm}
\omega_a (1) = + \frac{8}{9}\, \frac{1}{m^2} 
\nn \\
\omega_b(0) &=& + \frac{2}{9}\, \left(\frac{1}{m^2_s} - \frac{1}{m^2}\right)
\hspace{2.0cm}
\omega_b(1) =  - \frac{4}{27}\, \left(\frac{1}{m^2_s} - \frac{1}{m^2}\right)
\nn \\
\omega_c(0) &=& + \frac{1}{3}\, \frac{1}{m^2}  \hspace{2.0cm}
\omega_c(1) = - \frac{1}{9}\, \frac{1}{m^2} 
\nn \\
\omega^{(1)}_d(0) &=& - \frac{1}{9} \left(\frac{1}{m^2} + \frac{4}{m m_s}\right)
\hspace{2.0cm} 
\omega^{(1)}_d(1) = - \frac{1}{27} \left(\frac{1}{m^2} 
+ \frac{8}{m m_s}\right) \\
\omega^{(2)}_d(0) &=& + \frac{2}{9} \left(\frac{1}{m^2_s} 
+ \frac{1}{m m_s}\right)
\hspace{2.0cm} 
\omega^{(2)}_d(1) = - \frac{2}{27} \left(\frac{2}{m^2_s} 
- \frac{1}{m m_s}\right) 
\eea

\subsection{Contact spin-spin} 
\beq 
V^{\rm SS}_D (\vec{p},\vec{p}\,') =  \kappa_{ss} \, \omega_D(I)
\eta_D\, \exp \left[ -A_{D}\,p^{2} 
- B_{D}\,p\,'^{\,2} + C_{D}\, \vec{p} \cdot \vec{p}\,'\right] 
\label{vss-kn} 
\eeq

\vspace{0.25cm} 
\noindent 
\underline{Diagram (a):} 
\bea
A_{a} &=& \frac{2(1+\rho )^2+3g}{12\alpha ^2(1+\rho )^2} 
\hspace{1.0cm}
B_{a} = A_{a} 
\hspace{1.0cm} 
C_{a} = \frac{2(1+\rho )^2+3g}{6\alpha ^2(1+\rho )^2}
\nn \\
\eta_{a} &=& 1 \hspace{1.0cm} \omega_a (1) = 1/3 \hspace{1.0cm} \omega_a(0) = 0
\nn \\
\omega_a (1) &=& 1/3 \hspace{1.0cm} \omega_a(0) =0 \nn
\eea

\vspace{0.25cm} 
\noindent 
\underline{Diagram (b):} 
\bea
A_{b} &=&
\frac{(5g+3)\rho ^2+(-2g+6)\rho +(2g+3)}{6\alpha ^2(g+3)(1+\rho )^2}
\hspace{1.0cm}
B_{b} = \frac{(5g+3)\rho ^2+(10g+6)\rho +(5g+3)}{6\alpha ^2(g+3)(1+\rho )^2}
\nonumber\\
C_{b} &=&\frac{(1-g)\rho ^2+2\rho +(g+1)}{\alpha ^2(g+3)(1+\rho )^2}
\nonumber\\
\eta_{b}&=&\rho \,\left( \frac{6}{g+3}\right)^{3/2}\hspace{1.0cm}
\omega_b (1) = 1/3  \hspace{1.0cm} \omega_b (0) = 0 
\nn
\eea

\vspace{0.25cm} 
\noindent 
\underline{Diagram (c):} 
\bea
A_{c} &=&
\frac{(16g+3)\rho ^2+(\ 2g+6)\rho +(21g^2+22g+3)}{12\alpha
^2(7g+6)(1+\rho )^2} 
\nonumber\\
B_{c} &=&\frac{(64g+3)\rho ^2+(\ 62g+6)\rho +(21g^2+34g+3)}{12\alpha
^2(7g+6)(1+\rho )^2} 
\nonumber\\
C_{c} &=&\frac{(1-8g)\rho ^2+2\rho 
+\ (7g^2+8g+1)}{2\alpha ^2(7g+6)(1+\rho )^2} 
\nonumber \\
\eta_{c} &=& \left(\frac {12g}{7g+6}\right)^{3/2} \hspace{1.0cm}
\omega_c (1) = 1/18  \hspace{1.0cm} \omega_c(0) = 1/6 \nn
\eea

\vspace{0.25cm} 
\noindent 
\underline{Diagram (d):} 
\bea
A_{d} &=&
\frac{(20g^2+40g+3)\rho ^2+\ (4g^2+14g+6)\rho +\ (5g^2+10g+3)}{%
12\alpha ^2(2g+3)(g+2)(1+\rho )^2} \hspace{1.0cm} B_{d} = A_{d}
\nonumber\\
C_{d} &=&\frac{(1-4g^2-8g)\rho ^2+\ (2-4g^2-6g)\rho +\ (\ g^2+2g+1)}{\
2\alpha ^2(2g+3)(g+2)(1+\rho )^2 }\;\;.
\nonumber\\
\eta_{d}&=&\rho \,\left[\frac {12g}{\ (2g+3)(g+2)}\right]^{3/2}
\hspace{1.0cm} \omega_d (1) = 1/18 \hspace{1.0cm} \omega_d(0) = 1/6 \nn
\eea

\subsection{Contact spin-independent} 

\beq 
V^{\rm Con-SI}_D (\vec{p},\vec{p}\,') = 4\pi\alpha_s\, 
\omega_D(I) \, \exp  \exp \left[ - A_{D}\,p^{2} - B_D \,p\,'^{\,2}
+ C_D \, \vec{p} \cdot \vec{p}\,'\right], 
\label{vcons-kn} 
\eeq 
where the functions $A_{D}, \cdots$ are the same as for the spin-spin interaction
and

\bea
\omega_a (1) &=& + \frac{1}{9} \frac{1}{m^2} \hspace{1.0cm} \omega_a(0) = 0
\nn \\
\omega_b (1) &=& - \frac{1}{18} \frac{1}{m^2}\left(1 + \rho^2\right)
\hspace{1.0cm} \omega_a(0) = 0
\nn\\
\omega_c (1) &=& - \frac{1}{9} \frac{1}{m^2} \hspace{1.0cm} \omega_a(0) = 0
\nn \\
\omega_c (1) &=& - \frac{1}{18} \frac{1}{m^2} \left(1 + \rho^2\right)
\hspace{1.0cm} \omega_a(0) = 0
\eea

\subsection{Confinement} 

The confining interaction is taken to be a linearly rising potential, which
in momentum space is given as
\beq
V_{conf} = \frac{6\pi\sigma}{q^4},
\label{Vconf}
\eeq
where $\sigma$ is the string tension. Eq.~(\ref{Vconf}) includes a color factor 
of $3/4$. The effective $KN$ interaction is given by
\beq 
V^{\rm Conf}_D (\vec{p},\vec{p}\,') = 6\pi\sigma \,\omega_D(I)
\int_{0}^{\infty }\,du
\int_{u}^{\infty }\, ds \,
\eta_D(s)\, \exp \left[ - A_{D}(s)\,p^{2} 
- B_{D}(s)\,p\,'^{\,2} + C_{D}(s)\, \vec{p} \cdot \vec{p}\,'\right] ,
\label{vconf-kn} 
\eeq
where the functions $A_{D}(s), B_{D}(s), C_{D}(s)$ and $\eta_D$ are the same as 
for the Coulomb term, and the $\omega_D(I)$'s are given by:

\vspace{0.25cm} 
\noindent 

\bea
\omega_a (1) &=& + 4/9 \hspace{1.0cm} \omega_a(0) = 0  \nn \\
\omega_b (1) &=& - 4/9 \hspace{1.0cm} \omega_b(0) = 0 \nn \\
\omega_c (1) &=& - 4/9 \hspace{1.0cm} \omega_a(0) = 0 \nn \\
\omega_d (1) &=& + 4/9 \hspace{1.0cm} \omega_d(0) = 0 .
\eea

\newpage 

\begin{table}
\begin{center}
\caption{Vertex parameters used in the J\"ulich $KN$ model I
\protect\cite{Juel2}. Numbers in parentheses denote corresponding
values of the model discussed in the present paper, 
when different.}
\begin{tabular}{cccccc}
Process & Exch. part. & $M_r$ or $m_r \, ^{a)}$
 & $g_1g_2 /4 \pi \, ^{b)}$
 & $\Lambda_1 \, ^{c)}$  & $\Lambda_2 \, ^{c)}$   \\
        &         & [$MeV$]   & [$f_1/g_1$]
 & [$GeV$] & [$GeV$] \\
\hline
$K N \rightarrow KN$ & $\sigma$       & \phantom{1}600\phantom{.03}
 &\phantom{--4}1.300\phantom{[6.1]}
 & 1.7 & 1.5  \\
                       & & & \phantom{--4}(1.000)\phantom{[6.1]}
 &  & (1.2)  \\
                     & $\sigma_{rep}$ & 1200\phantom{.03}
 & --40\phantom{.000}\phantom{[6.1]}
 & 1.5 & 1.5  \\
                     & & & (--)
& (--) & (--)  \\
                     & $a_0$ & 980\phantom{.03} & -- & -- & --  \\
                     & & & (2.600)\phantom{[6.1]}
 & (1.5) & (1.5)  \\
                     & $\omega$       & \phantom{1}782.6\phantom{3}
 &\phantom{--4}2.318 [0]\phantom{.1}
 & 1.5 & 1.5  \\
                     & $\rho$         & \phantom{1}769\phantom{.03}
 &\phantom{--4}0.773[6.1]
 & 1.4 & 1.6  \\
                     & $\Lambda$      & 1116\phantom{.03}
 &\phantom{--4}0.905\phantom{[6.1]}
 & 4.1 & 4.1  \\
                     & $\Sigma$       & 1193\phantom{.03}
 &\phantom{--4}0.031\phantom{[6.1]}
 & 4.1 & 4.1  \\
                     & $Y^*$          & 1385\phantom{.03}
 & \phantom{--4}0.037\phantom{[6.1]}
 & 1.8 & 1.8  \\
$K N \rightarrow K^* N$ & $\pi$          & \phantom{1}138.03
 &\phantom{--4}3.197\phantom{[6.1]}
 & 1.3 & 0.8  \\
                        & $\rho$         & \phantom{1}769\phantom{.03}
 &\phantom{--4}0.773[6.1]
 & 1.4 & 1.0  \\
$K N \rightarrow K^* \Delta$ & $\pi$          & \phantom{1}138.03
 &\phantom{--4}0.506\phantom{[6.1]}
 & 1.2 & 0.8  \\
                          & $\rho$         & \phantom{1}769\phantom{.03}
 &\phantom{--4}4.839\phantom{[6.1]}
 & 1.3 & 1.0  \\
$KN \rightarrow K \Delta$ & $\rho$         & \phantom{1}769\phantom{.03}
 &\phantom{--4}4.839\phantom{[6.1]}
 & 1.3 & 1.6
\end{tabular}
\vskip0.5cm
\noindent
\begin{flushleft}
$^{a)}$ Mass of exchanged particle. \\
$^{b)}$ Product of coupling constants [ratio of tensor to vector
        coupling]. \\
$^{c)}$ Cutoff mass. \\
\end{flushleft}
\end{center}
\label{PARAM}
\end{table}

\newpage

\begin{figure}
\caption{Meson-exchange contributions to $KN$ scattering in the J\"ulich model
[1,2]. Diagrams (a) and (b) define the model of Ref. [1] and diagram (c) is
the correlated $2\pi$ exchange calculated in Ref. [2], which was parametrized 
by diagram (a) in Ref. [1].}
\label{Juel_mod}
\end{figure}

\begin{figure}
\caption{The four quark-interchange kaon-nucleon scattering diagrams.}
\label{FT-KN}
\end{figure}

\begin{figure}
\caption{
$KN$ phase shifts. The solid line are the result from the full 
quark-model calcuation described in the text. 
Experimental phase shifts are
taken from Ref.~\protect\cite{Exp1} (open circles),
Ref.~\protect\cite{Exp2} (open squares), and
Ref.~\protect\cite{Exp3} (filled circles and pluses)
.}
\label{phasqf} 
\end{figure}

\begin{figure}
\caption{
$KN$ phase shifts resulting from various contributions of the
quark model, cf. Appendix A: 
Coulomb (dash-dotted line);
Spin-orbit (pluses);
Confinement (short dashed line);
Contact spin-spin (long dashed line);
Contact constant (solid line).
}
\label{phasq} 
\end{figure}

\begin{figure}
\caption{
$KN$ phase shifts. The dash-dotted line are the phase shifts
of the original J\"ulich model I from Ref.~\protect\cite{Juel2}.
The short dashed line shows results where the phenomenological
$\sigma_{rep}$ in the J\"ulich model is replaced by the quark-model 
contribution. Adding the $a_0$-exchange contribution yields
the long dashed line. The solid line is obtained after refitting
the parameters of the $\sigma$. Same description of experimental 
phase shifts as in Fig.~\ref{phasqf}.}
\label{phases} 
\end{figure}

\newpage 

\begin{figure}
\centerline{\epsfxsize=12.0cm\epsfysize=15.0cm\epsfbox{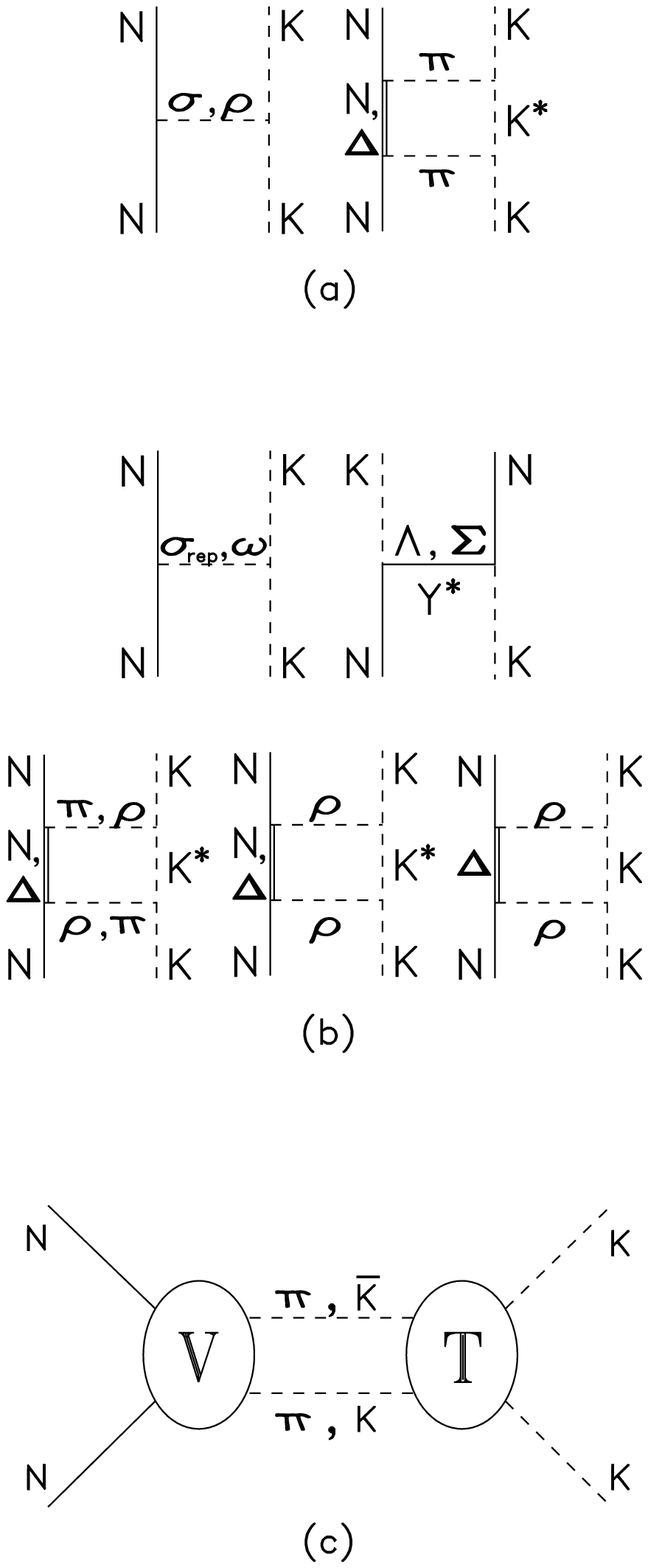} }
\end{figure}

\vskip 4cm

\center{Fig. 1}

\newpage

\begin{figure}
\centerline{
{\epsfxsize=10.0cm\epsfbox{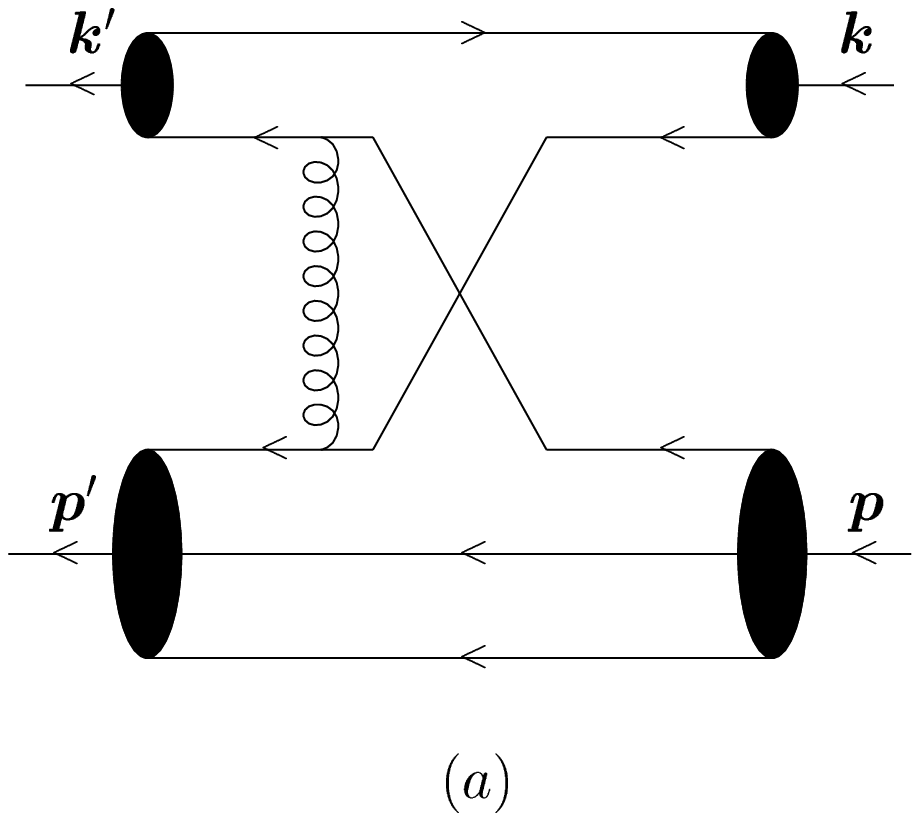}}\hspace{-3.0cm}
{\epsfxsize=10.0cm\epsfbox{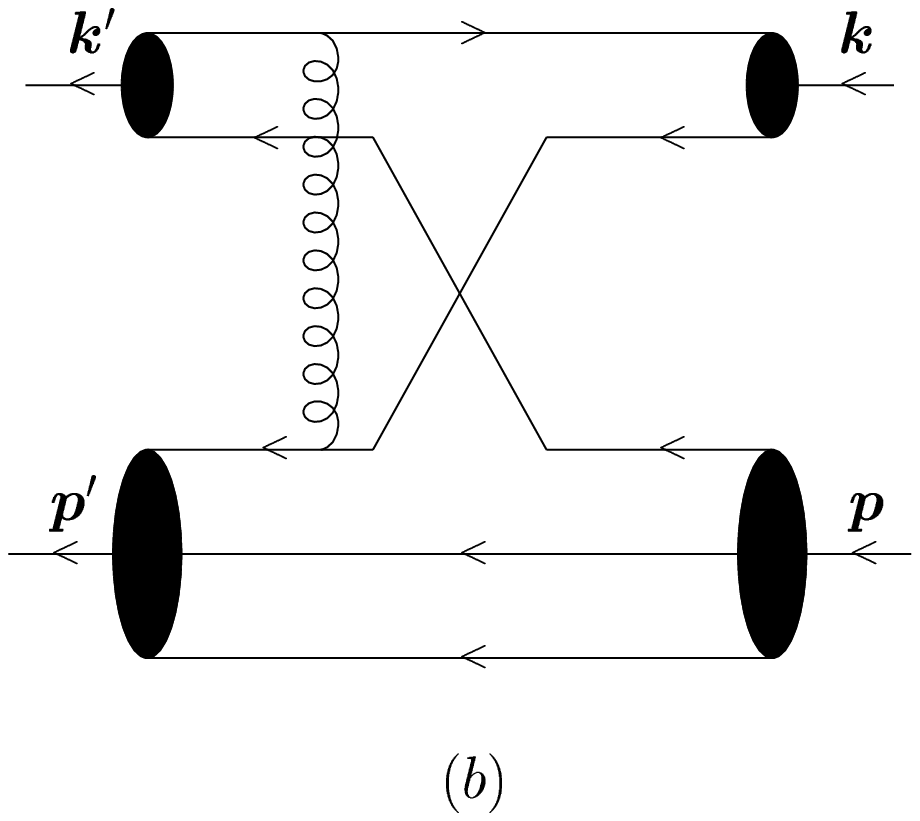}}
}
\vspace{-9.0cm}
\centerline{
{\epsfxsize=10.0cm\epsfbox{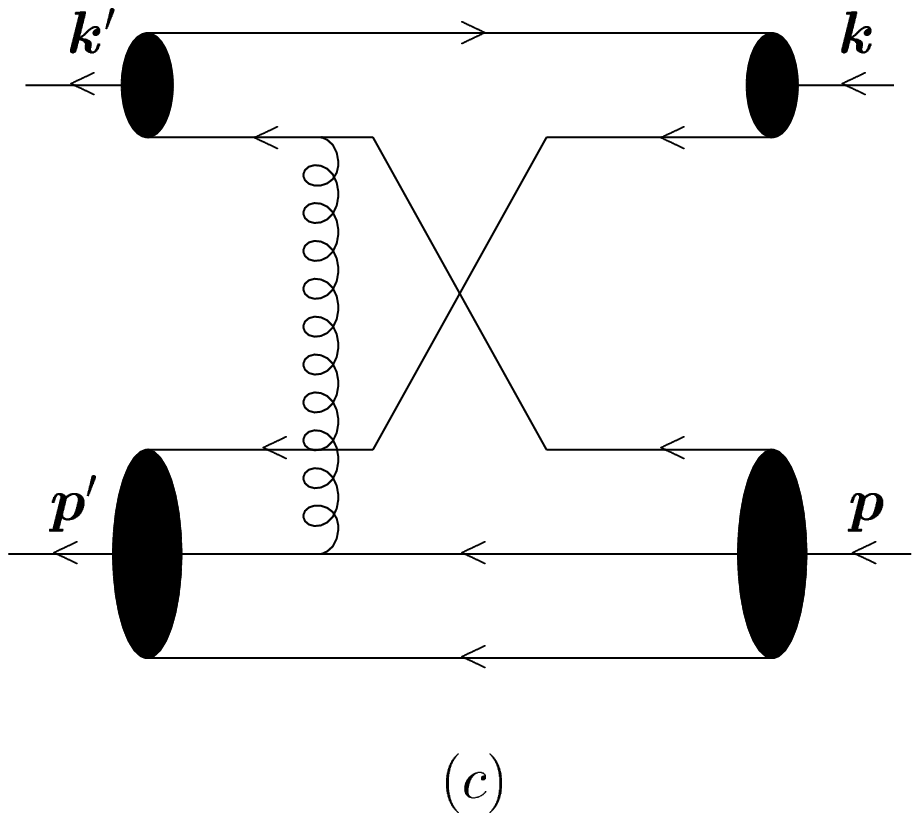}}\hspace{-3.0cm}  
{\epsfxsize=10.0cm\epsfbox{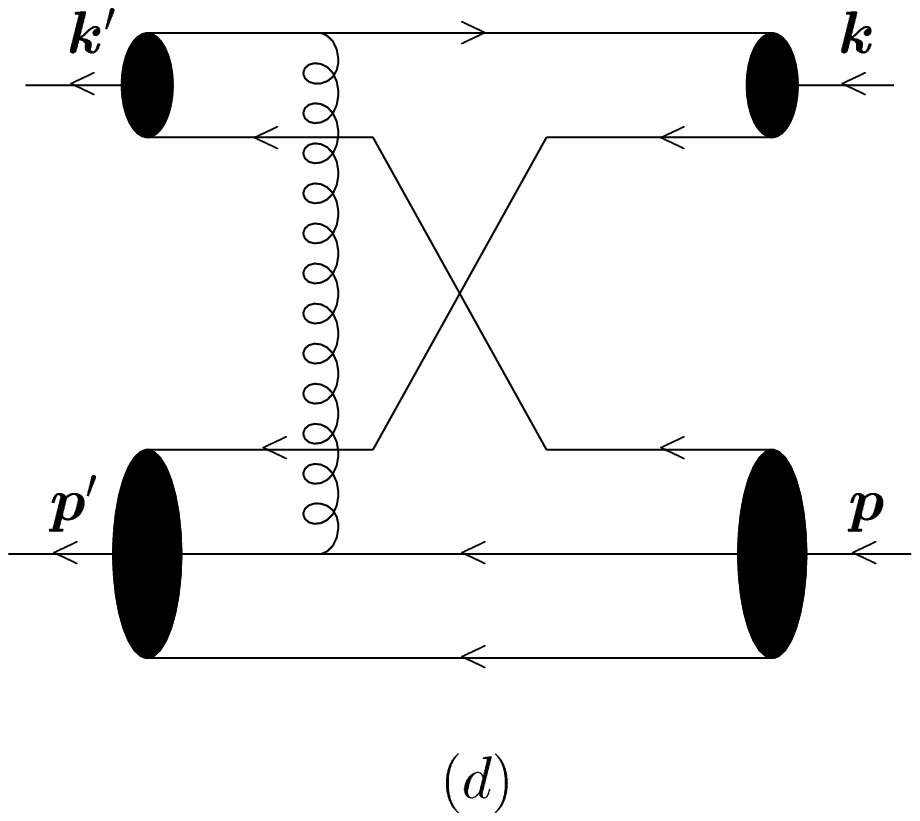}} 
}
\vspace{-8.0cm}
\end{figure}

\vskip 4cm

\center{Fig. 2}

\newpage

\vglue 1cm
\begin{center}
\begin{figure}
\epsfig{file=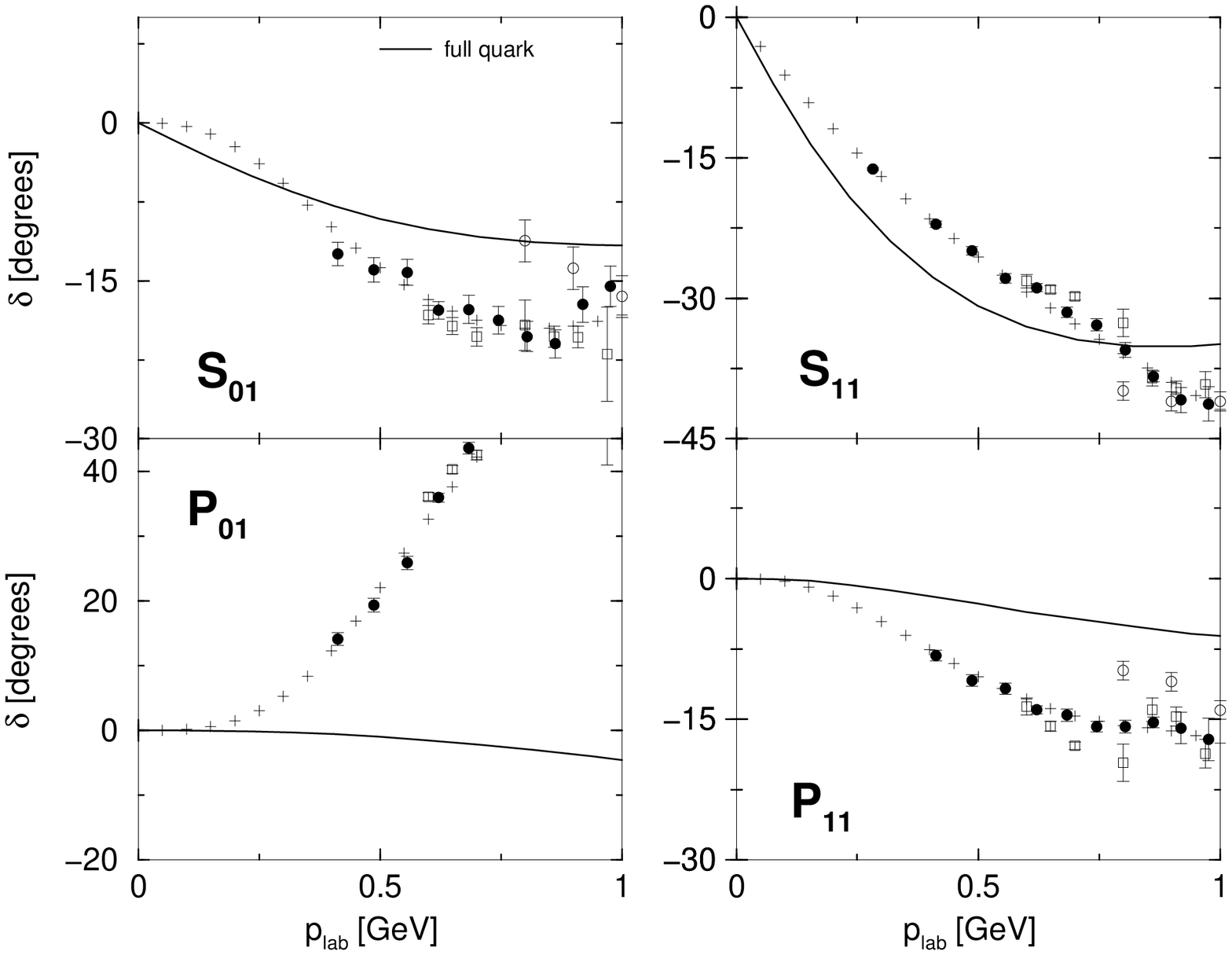, width=15.0cm}
\end{figure}
\end{center}

\vskip 4cm

\center{Fig. 3}

\newpage

\vglue 1cm
\begin{center}
\begin{figure}
\epsfig{file=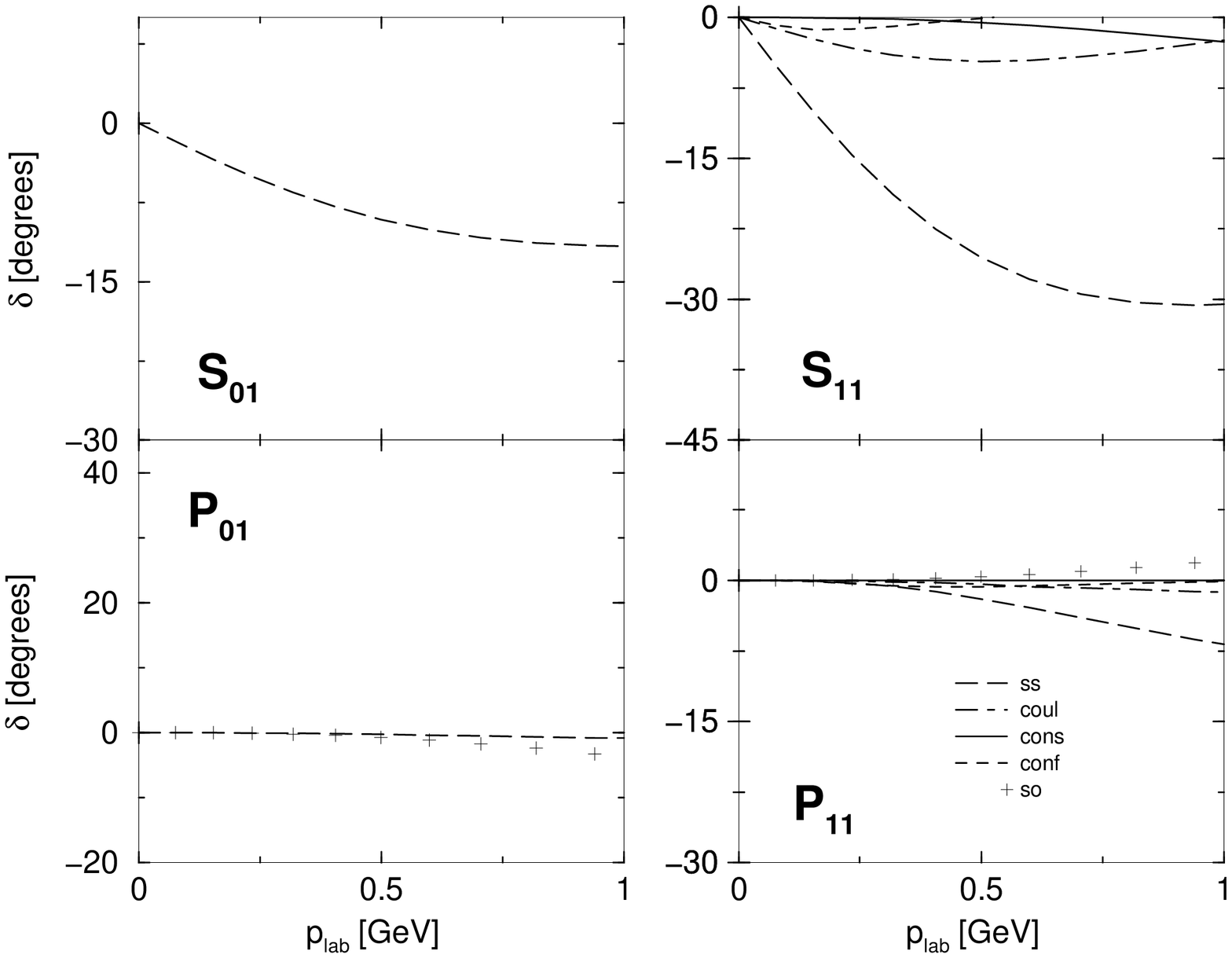, width=15.0cm}
\end{figure}
\end{center}

\vskip 4cm

\center{Fig. 4}

\newpage

\vglue 1cm
\begin{center}
\begin{figure}
\epsfig{file=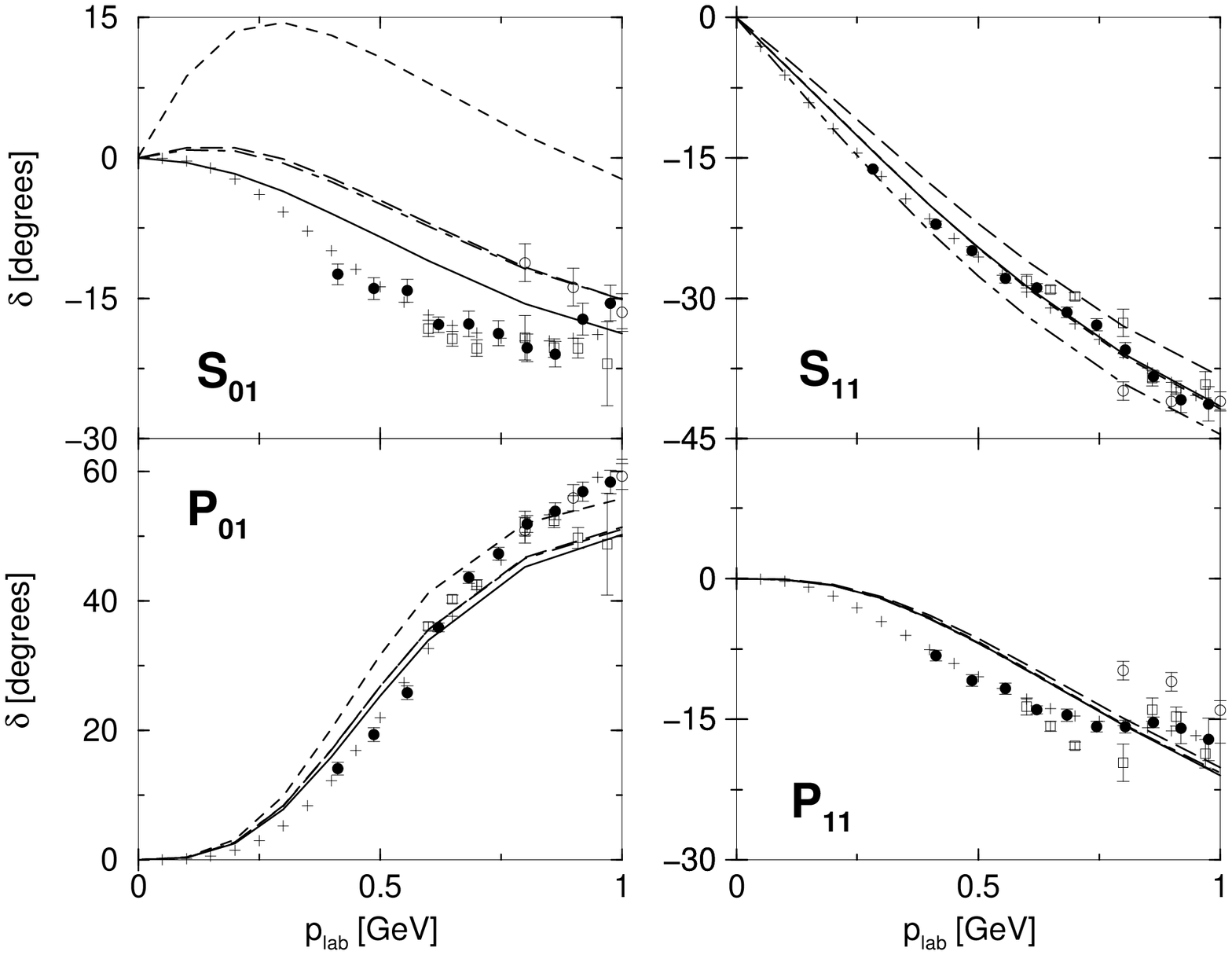, width=15.0cm}
\end{figure}
\end{center}

\vskip 4cm

\center{Fig. 5}

\newpage

\vglue 1cm
\begin{center}
\begin{figure}
\epsfig{file=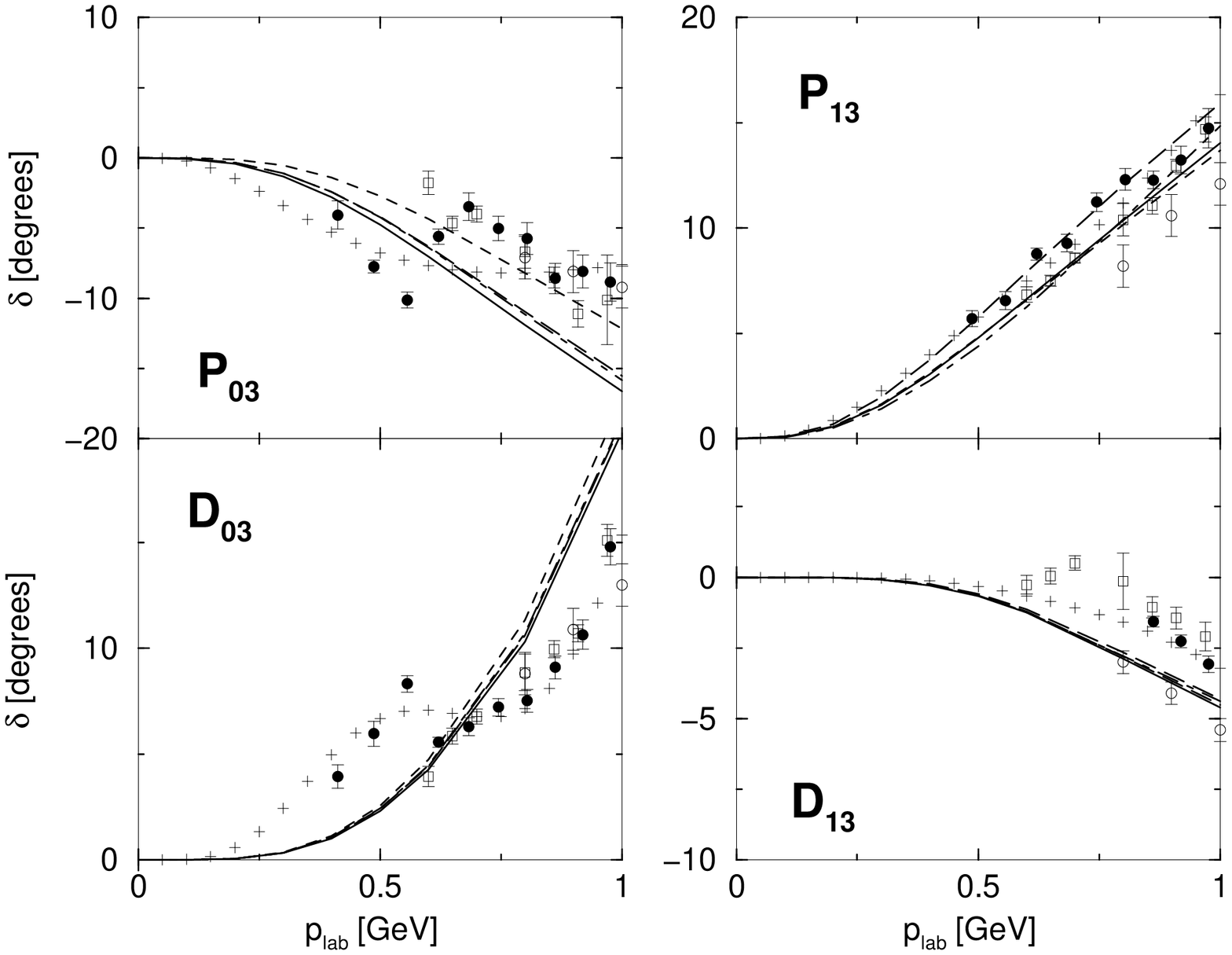, width=15.0cm}
\end{figure}
\end{center}

\vskip 4cm

\center{Fig. 5, cont.}

\newpage

\vglue 1cm
\begin{center}
\begin{figure}
\epsfig{file=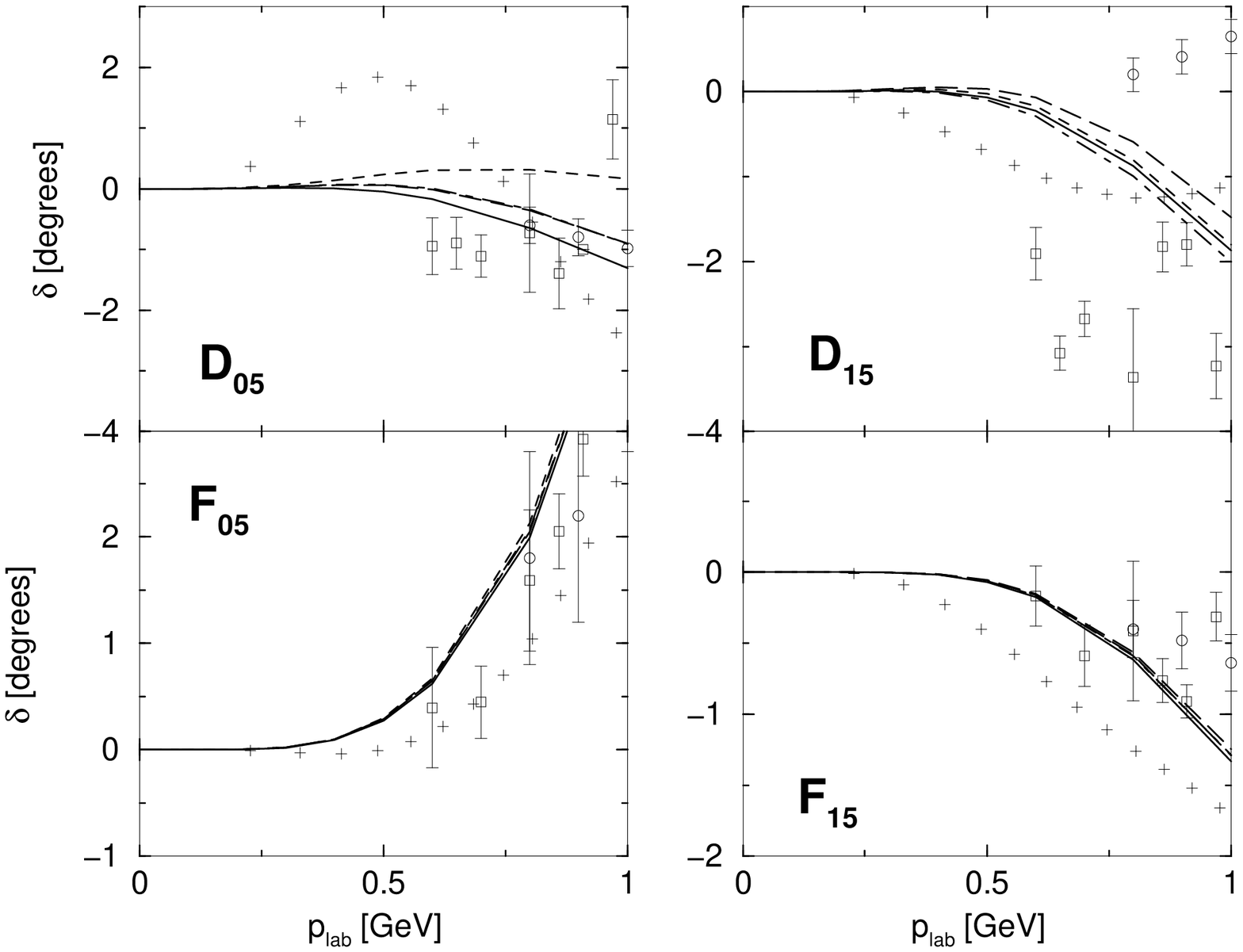, width=15.0cm}
\end{figure}
\end{center}

\vskip 4cm

\center{Fig. 5, cont.}

\newpage

\vglue 1cm
\begin{center}
\begin{figure}
\epsfig{file=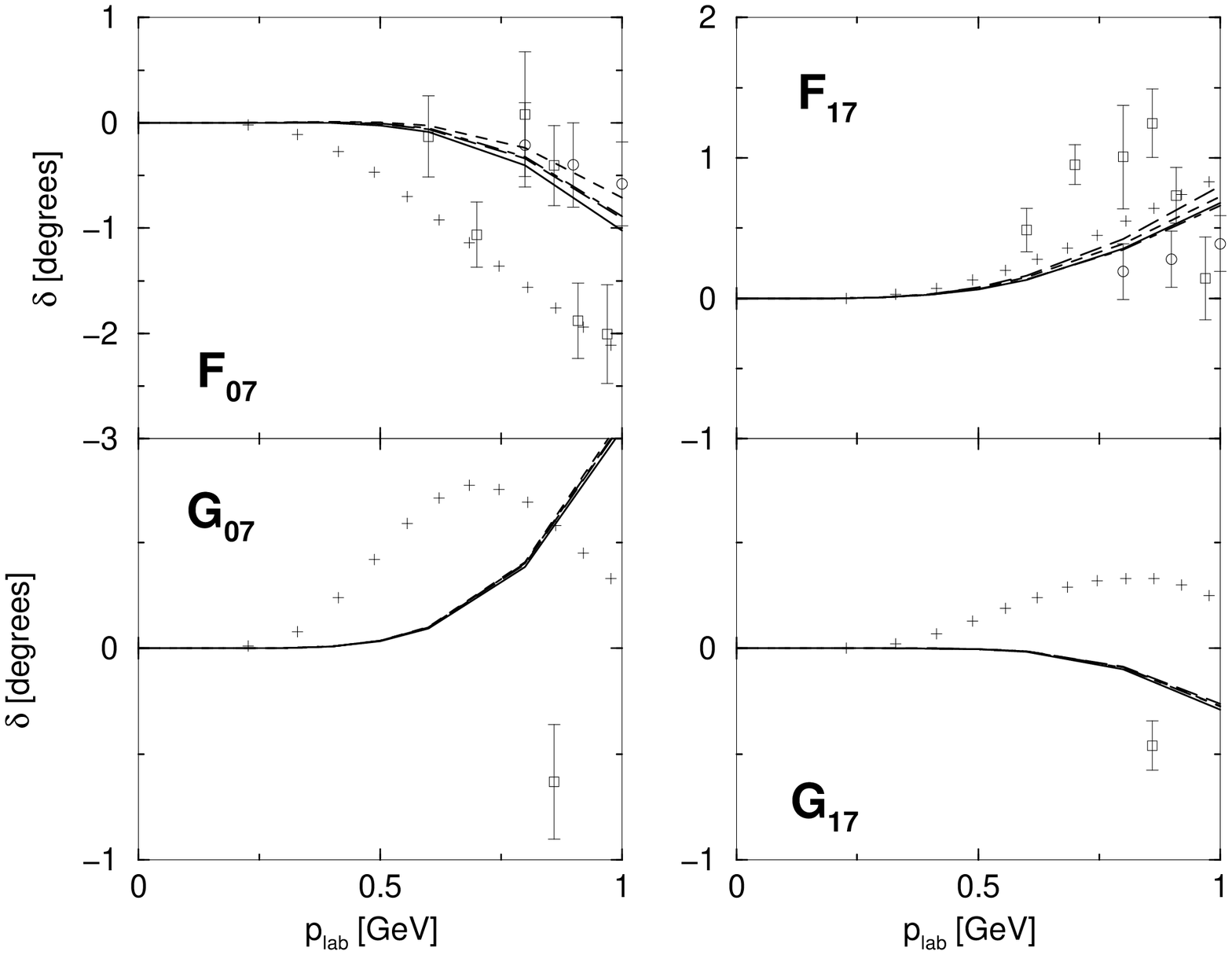, width=15.0cm}
\end{figure}
\end{center}

\vskip 4cm

\center{Fig. 5, cont.}

\end{document}